\documentstyle[11pt]{article}
\font\tenbf=cmbx10
\font\tenrm=cmr10
\font\tenit=cmti10
\font\elevenbf=cmbx10 scaled\magstep 1
\font\elevenrm=cmr10 scaled\magstep 1
\font\elevenit=cmti10 scaled\magstep 1

\renewenvironment{thebibliography}[1]
 { \elevenrm
   \begin{list}{\arabic{enumi}.}
    {\usecounter{enumi} \setlength{\parsep}{0pt}
     \setlength{\itemsep}{1pt} \settowidth{\labelwidth}{#1.}
     \sloppy
    }}{\end{list}}

\def\ee{e^+e^-}
\def\to{\rightarrow}
\def\bbar{{\bar{b}}}
\def\bb{{b\bar{b}}}
\def\tbar{{\bar{t}}}
\def\tt{t\bar{t}}
\def\degree{^{\circ}}

\def\GeV{{\rm GeV}}
\def\cM{{\cal M}}
\def\cR{{\cal R}}
\def\as{\alpha_s}

\def\beq{\begin{equation}}
\def\eeq{\end{equation}}

\begin{document}
\vspace*{-1cm}
\begin{flushright}
UCD-92-28 \\
 DTP/92/82   \\
     November 1992 \\
\end{flushright}
\vspace*{.5cm}
\begin{center}{{\tenbf TOP WIDTH EFFECTS IN SOFT GLUON RADIATION\footnote{
Work supported in part by the Texas National Research Laboratory
Commission and the United Kingdom Science and Engineering Research Council.
}$^,$\footnote{Presented by L.H. Orr at the 1992 Meeting of the
Division of Particles and Fields, Fermilab, November 10-14, 1992.}
\\}
\vglue 0.6cm
{\tenrm LYNNE H. ORR \\}
\baselineskip=12pt
{\tenit Department of Physics, University of California\\}
\baselineskip=11pt
{\tenit Davis, CA 95616, USA\\}
\vglue 0.25cm
{\tenrm YU.L. DOKSHITZER \\}
{\tenit Department of Theoretical Physics, University of Lund \\
S\"olvegatan 14A, S-22362 Lund, Sweden \\}
\vglue 0.15cm
{\tenrm and\\}
\vglue 0.15cm
{\tenrm V.A. KHOZE and W.J. STIRLING \\}
{\tenit Department of Physics, University of Durham \\
Durham DH1 3LE, England\\}
\vglue 0.4cm
{\tenrm ABSTRACT}}
\end{center}
{\rightskip=2pc
 \leftskip=2pc
 \tenrm\baselineskip=12pt
 \noindent
Soft gluons radiated in top quark production and decay can interfere in
a way that is sensitive to the top width.  We show how the width affects
the gluon distribution in $e^+e^- \rightarrow t {\bar t}$ and discuss prospects
for measuring $\Gamma$ from gluons radiated
near $\tt$ threshold.
\vglue 0.3cm}
{\elevenbf\noindent 1. Soft Gluon Radiation in $\ee \to \tt$}
\vglue 0.2cm
\baselineskip=13pt
\elevenrm
Because the top quark is so heavy that it can decay to a real $W$ and a $b$,
it has a very large width: for
large $m_t$, $\Gamma(t\rightarrow W b) \approx (175\ {\rm MeV}) (m_t/m_W)^3$.
Widths in the GeV range can give rise to interesting effects involving the
interplay between the strong and weak interactions.  For example, if top
is heavy enough, it can decay before forming bound states, and
there is not much resonant structure at the $\tt$ threshold, making it
difficult
to measure $\Gamma$.
In this talk we
consider the effect of the top width on soft gluon radiation in
$\ee \to \tt$, at arbitrary collision energies$^{[1]}$ and near
$\tt$ threshold.$^{[2]}$
For complete discussions see Refs. [1] and [2].

Consider a gluon emitted in a $\tt$ event at an $\ee$ collider.  Because of
the top decays, the gluon can be radiated by
the $t, \tbar, b$, or $\bbar$.  In the limit
of soft gluons, the matrix element $\cM$ factorizes and can be written as
a product of the zeroth-order matrix element (with no gluon) and
a term associated with the gluon emission.  Schematically,
we have $\cM \sim \cM^{(0)} J\cdot\epsilon$, where $J^{\mu}$ and
$\epsilon_{\mu}$ are the gluon current and polarization, respectively.
We can then define a gluon emission probability density, which is just the
differential cross section for radiating a gluon normalized to the
zeroth-order cross section.  It is given by
\beq
dN \equiv 1/\sigma_0 d\sigma_g =
 \frac{d\omega}{\omega}\frac{d\Omega}{4\pi}\>\frac{C_F\as}{\pi}\>\cR\>,
\eeq
where $\omega$ and $\Omega$ denote the gluon energy and solid angle.
$\cR$ is obtained by integrating the absolute square of the current over the
virtualities
of the $t$ and $\tbar$.

The important point is that the current can be decomposed in a gauge-invariant
way into terms corresponding to order $\as$ corrections to $\tt$
production, to $t$ decay, and to $\tbar$ decay.  Therefore in $\cR$ we can
unambiguously identify each contribution:
(production)$^2$, ($t$ or $\tbar$ decay)$^2$, production--decay interference
and decay--decay interference.
The production and decay squared terms are independent
of the width, but both interference terms have $\Gamma$ dependence, including
an overall factor of $\Gamma^2$.

\begin{figure}
\vspace{14cm}
\vspace{7cm}
\hspace{-3.2cm}
\vspace{-14.95cm}
\caption{Soft gluon distribution in $e^+e^-\to t \tbar$ for
c.m. energy 1 TeV, $m_t=140\ \GeV$,
 $\omega = 5\ \GeV$ and $\phi=0\degree$.  $\theta$ is the $t$--$g$ angle;
the $t$ and $\bbar$
are at $0\degree$ and the $\tbar$ and $b$ are at $180\degree$. }
\vspace{-.23cm}
\end{figure}

\vglue 0.3cm
{\elevenbf\noindent 2. Width Effects at High Energies}
\vglue 0.2cm

At high collision energies, the top width dependence arises from
production--decay interference; the decay--decay interference is negligible.
The production--decay interference is largest for large $b$--$t$ angular
separations, and is destructive, so that the effect of the width is to
suppress the gluon radiation.

This is illustrated in Fig. 1, where we
show the gluon emission probability as a function of the angle $\theta$ of
the gluon with respect to the top quark.  We vary the top width
and take $m_t=140\ \GeV$, $\omega=5\ \GeV$, and
center-of-mass energy 1 TeV.  The $t$ and $\tbar$ are produced back-to-back,
and we have chosen a configuration in which the
$t (\tbar)$ decays to a backward $b(\bbar)$.
We see that for the SM case ($\Gamma=0.7\ \GeV$) the peaks are suppressed
compared to the case with no interference ($\Gamma=0$) and as
the width increases the peaks disappear altogether.

Now,  energetic
top quarks do not often decay to backward $b$'s.  If we take a
slightly more likely $b$--$t$ angle such as $90\degree$, we obtain
similar sensitivity.
However, in the most probable configuration --- $t$ and $b$ collinear ---
there is almost no interference and therefore no sensitivity to the top
width.
Thus at high energies, the production--decay interference can be
substantial, but the most sensitive configurations are the least likely to
occur.

\vglue 0.3cm
{\elevenbf\noindent 3. Width Effects Near $\tt$ Threshold}
\vglue 0.2cm

At lower energies, near the $\tt$ threshold,  the total cross section is higher
and the $t$'s are produced nearly at rest, so that the relative orientations
of the $t$ and $b$ momenta are irrelevant.  We might expect, then, that
top width effects could be more pronounced.  On the other hand,  if
the $t$'s are nearly at rest and only the $b$'s can radiate,
it is not obvious that the top width enters at all.  Naively, one would
expect the $b$'s to radiate as if they were produced directly and the
$t$'s never existed.

That the top width {\it does} influence the radiation from the $b$'s can
be understood by considering the following extreme cases.  As
$\Gamma \to \infty$,
the top lifetime becomes very short, the $b$ and $\bbar$ appear
almost instantaneously, and they radiate coherently, as
though produced directly.  In particular, gluons from the $b$ and $\bbar$
interfere.  In the other extreme, for $\Gamma \to 0$,
top has a long lifetime and the $b$ and $\bbar$ appear at very different
times and therefore radiate independently, with no interference.
Clearly, the top width controls the interference between gluons
emitted by the $b$ and $\bbar$.

The situation for finite width is between the two extremes.
Let $v$ be the $b$
(or $\bbar$) velocity, $\theta_{1(2)}$ be the angle between the $b$ ($\bbar$)
and the gluon, and $\theta_{12}$ the angle between the $b$ and $\bbar$.  Then
\beq
\cR = \frac{v^2\sin^2\theta_1}{(1-v\cos\theta_1)^2}\> +\>
\frac{v^2\sin^2\theta_2}{(1-v\cos\theta_2)^2}\>
+\>
2 \chi\,
\frac{v^2(\cos\theta_1\cos\theta_2-\cos\theta_{12})}
{(1-v\cos\theta_1)(1-v\cos\theta_2)},
\eeq
where $\chi \equiv { \Gamma^2 \over \Gamma^2 + \omega^2}$.
The interference is the term proportional to $\chi$.  Note that
$0 \leq\chi\leq1$ and  $\chi=0$ for
$\Gamma=0$ (independent emission) and $\chi=1$ for $\Gamma=\infty$
(coherent emission).
Thus a finite top width suppresses the interference compared to the naive
expectation of full coherent emission.  And from the form of $\chi$ we
see that the radiation pattern exhibits maximum sensitivity to $\Gamma$
when $\Gamma$ is comparable to the gluon energy $\omega$.

\begin{figure}
\vspace{14cm}
\vspace{7cm}
\hspace{-3.2cm}
\vspace{-15.65cm}
\caption{Soft gluon distribution in $e^+e^-\to t \tbar$ for
$m_t=140\ \GeV$,
near $\tt$ threshold, with gluon perpendicular to $\bb$ plane;
$\theta_{12}$ is the $b$--$\bbar$ angle.}
\vspace{-.28cm}
\end{figure}

The width effects are discussed in detail in Ref. 2;
here we give two examples.  First consider gluons emitted
perpendicular to the $\bb$ plane.  Then $\theta_1=\theta_2=\pi/2$; $\cR$ is
simply proportional to $1-\chi\cos\theta_{12}$ and $\chi$ regulates the
$\theta_{12}$ dependence --- see Fig. 2.
Now for a 5 GeV gluon, a 140 GeV top quark with $\Gamma=0.7\ \GeV$
has $\chi\approx 0.02$, which means the distribution is much closer to the
independent
emission case ($\chi=0$) than the coherent case we would naively expect --
the interference is almost completely absent.
If we could detect
1 GeV gluons, we would have $\chi\approx 0.3$ and the distribution would be
very sensitive to the width.  (Conversely, we would get the same sensitivity
for more energetic gluons if the width were larger:  $\chi\approx 0.3$
with gluon energy 5 GeV corresponds to $\Gamma\approx 3\ \GeV$.)

As a second example we show in Fig. 3 the gluon distribution integrated over
the gluon
solid angle (and in a slight nod to reality, integrated over
gluon energies
from 5 to 10 GeV).  For  independent emission the
radiation probability does not depend on the angle between the $b$ and $\bbar$,
but in the coherent case the interference is destructive for small
$\theta_{12}$ and
constructive for large $\theta_{12}$.
Again, we see that the 140 GeV case is much closer to
the independent than the expected coherent case, and that as the width
reaches the few GeV range we become increasingly sensitive.

\begin{figure}
\vspace{14cm}
\vspace{7cm}
\hspace{-3.2cm}
\vspace{-15.5cm}
\caption{Soft gluon emission probablilty near $\tt$ threshold
($m_t=140\ \GeV$)
integrated over gluon angles and energies
from 5 to 10 GeV.}
\vspace{-.23cm}
\end{figure}

\vglue 0.3cm
{\elevenbf\noindent 4. Discussion}
\vglue 0.2cm

Is looking at soft gluon radiation a useful method for measuring the top
width, an alternative to studying the threshold structure$^{[3]}$
of the lowest order cross section?
Each method has its disadvantages:  The threshold structure is subject to
large uncertainties due
to beam energy spread; the soft gluon radiation
is not, but it is a higher order process with a lower event rate.
The two methods should be considered complementary, because
the threshold cross section loses sensitivity with increasing width, but as we
have seen,
the gluon radiation pattern becomes {\it more} sensitive at
larger $\Gamma$ for accessible gluon energies.
The bottom line is that for most of the expected top
mass range, the threshold structure is probably better, but if
$m_t$ and $\Gamma$ are large, examining  soft gluons may be more useful.

In summary, we have seen that the top quark's large width gives rise to
new effects from the interplay between the strong and weak interactions,
and that the top width affects the distributions of soft gluons radiated
in top events.  At high collision energies, production--decay interference
can suppress gluon radiation.  Near
the $\tt$ threshold, the effect of the width is to suppress the
{\it interference} between gluons radiated by the $b$ and $\bbar$,
in contrast to the expectation of coherent radiation from the $\bb$ pair.
Finally, if the width and gluon energy are comparable, the radiation
pattern is quite sensitive to the value of $\Gamma$.

\vglue 0.3cm
{\elevenbf\noindent References \hfil}


\begin{thebibliography}{9}
\bibitem{kos} V.A. Khoze, W.J. Stirling and L.H. Orr, {\elevenit Nucl. Phys.}
{\elevenbf B378} (1992) 413.
\bibitem{dkos} Yu. L. Dokshitzer, V.A. Khoze, L.H. Orr, and W.J. Stirling,
in preparation.
\bibitem{fujii} K. Fujii, KEK preprint 92-6 (1992) and references therein.
\end{thebibliography}
\end{document}